\documentclass[letterpaper, 10 pt, conference]{IEEEtran}

\usepackage{afterpage}
\usepackage{algorithm}
\usepackage{algpseudocode}
\usepackage{pifont}
\usepackage{wrapfig}
\usepackage{amsmath,amssymb}
\usepackage{bm}
\usepackage[authormarkup=none,final]{changes}
\usepackage{epigraph} 
\usepackage{booktabs}
\usepackage{float}
\usepackage{graphicx}
\graphicspath{{./Figures/}}
\usepackage{lettrine}
\usepackage{mathtools}
\usepackage{mathrsfs}
\usepackage{multirow}
\usepackage{hyperref}
\usepackage[numbers]{natbib}
\usepackage[percent]{overpic}
\usepackage{comment}

\usepackage{pgfplots}
\usepackage[T1]{fontenc}
\usepackage{upgreek}
\usepackage{xcolor}
\usepackage{xfrac}
\usepackage{siunitx}
\def\BState{\State\hskip-\ALG@thistlm}
\usepackage{enumitem}
\usepackage{blindtext}
\SetLabelAlign{parright}{\smash{\parbox[t]{\labelwidth}{\raggedleft#1}}}
\usepackage{amsthm}

\usepackage[top=54pt,bottom=54pt,left=54pt,right=54pt]{geometry}
\usepackage{epstopdf}
\AppendGraphicsExtensions{.tif}

\newcommand{\matrixstyle}[1]{\bm{\mathrm{#1}}}   
\newcommand{\vectorstyle}[1]{\bm{#1}}            

\IEEEoverridecommandlockouts

\newtheorem{assumption}{\textbf{Assumption}}

\pgfplotsset{compat=1.18}
\begin{document}

\title{Aeroelastically Constrained Optimal Path Planning of a Fixed-Wing Airborne Wind Energy System}
\author{Omid Heydarnia\textsuperscript{*}, Jolan Wauters, Tom Lefebvre, Joris Degroote, and Guillaume Crevecoeur\vspace*{-20pt}
    \thanks{\textsuperscript{*}Corresponding author.}
    \thanks{Omid Heydarnia, Tom Lefebvre, Joris Degroote, and Guillaume Crevecoeur are with the Department of Electromechanical, Systems and Metal Engineering, Ghent University, B-9052, Ghent, Belgium. E-mail: omid.heydarnia@ugent.be.}
    \thanks{Jolan Wauters is with the Department of Mechanical Engineering, KU Leuven, B-8200, Bruges, Belgium.}
    \thanks{Omid Heydarnia, Tom Lefebvre, Joris Degroote, and Guillaume Crevecoeur are members of core-lab MIRO, Flanders Make, Belgium.}%
}

\maketitle
\thispagestyle{empty}
\pagestyle{empty}

\begin{abstract}

    Multi-megawatt fixed-wing airborne wind energy systems (AWES) use an aircraft with large, slender wings that deform over the pumping cycle. To harvest maximum energy, the cycle is usually designed by means of dedicated reference trajectory optimization routines. It is well known that such routines rely heavily on the system model and that close alignment between the model and the physical system is required for the results to hold in practice. Nevertheless, the wing is usually treated as rigid and structural limits are respected indirectly through upper bounding of states and inputs. In this work, we directly enforce bounds on the elastic deflection and twist of the wing in the power-maximizing periodic optimal control problem. To that end, we model the wing by root-clamped modes, loaded by strip-theory aerodynamics and by the gravitational and inertial loads of the manoeuvring aircraft. To keep the problem tractable, the modal dynamics are replaced by their quasi-static equilibrium, which enters every discretization interval as algebraic constraints.
    For a multi-megawatt reference aircraft, the elastic twist near the wing tips reaches $1.03^\circ$ along the optimal rigid-wing trajectory. With a twist limit of $0.6^\circ$, the optimal
    trajectory delivers $31\,\%$ less mean power than the rigid-wing one, mainly through a lower tether tension during reel-out. Rigid-wing trajectory optimization can thus overestimate the power that the wing limits permit. Forward simulations that retain the modal dynamics demonstrate that the quasi-static deflection and twist deviate from the dynamic ones by less than $2\,\%$ of their limits. This work shows that elastic deformation can be accounted for in the trajectory optimization and that, depending on the wing limits, it can strongly reduce the achievable power.

\end{abstract}

\section{INTRODUCTION}\label{sec:intro}

\begin{figure}[t!]
    \centering
    \includegraphics[width=\columnwidth]{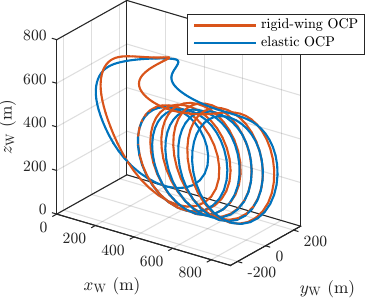}
    \caption{Trajectories from the optimal control problem (OCP) for the rigid wing and for the elastic wing with a twist limit of $0.6^\circ$.}
    \label{fig:traj}
\end{figure}

Airborne wind energy (AWE) systems replace the tower and blades of a conventional horizontal-axis wind turbine with a tethered fixed-wing aircraft that flies crosswind patterns. During a full pumping cycle, the aircraft pulls the tether from a ground-based winch at high tension during a traction phase and is reeled back in at low tension during a retraction phase, as shown in Fig.~\ref{fig:traj}. The cycle is designed so that the generator harvests net energy. Reference designs for multi-megawatt systems, such as the MegAWES aircraft with its \SI{42.5}{m} wing span~\cite{eijkelhof2022six}, rely on large, slender wings. In practice, these wings deform under the aerodynamic and inertial loads of crosswind flight. 

Trajectory optimization is the standard tool to find power-optimal flight paths of AWE systems. Periodic optimal control problems are formulated and solved with direct methods in frameworks such as AWEbox~\cite{de2023awebox} and Opti-MegAWES~\cite{heydarnia2025optimal}.
The optimized trajectories then serve as references for tracking controllers, for instance Model Predictive Control (MPC) evaluated in Large-Eddy-Simulation (LES) turbulence~\cite{crismer2026large} or a Linear Quadratic Regulator (LQR) of the coupled winch, tether and aircraft dynamics~\cite{heydarnia2026airborne}. The listed frameworks model the aircraft as a rigid body. Structural loads are limited only indirectly, through bounds on tether force, accelerations or aerodynamic angles, which do not guarantee that the wing stays within its limits.

The aeroelastic behaviour of the wings has mostly been addressed by post-processing analysis. Wijnja et al.~\cite{wijnja2018aeroelastic} carried out an aeroelastic analysis of a large airborne wind turbine, and high-fidelity fluid-structure interaction simulations quantified the wing deformation of the MegAWES reference aircraft in crosswind flight along an optimal trajectory, assuming the aircraft to be rigid in the trajectory optimization~\cite{pynaert2022high,pynaert2023wing}. Fasel et al.~\cite{fasel2019reduced} derived a reduced-order dynamic model that couples the aeroelasticity of a morphing AWE wing with the flight dynamics of the aircraft. These studies evaluate the wing along a given flight path, and the high-fidelity models in particular are far too expensive to be embedded in an optimizer. Closest to this work, Fasel et al.~\cite{fasel2021concurrent} embedded a quasi-steady reduced-order aeroelastic wing model in an evolutionary optimization of a morphing AWE wing, simultaneously optimizing control gains and a parametrized circular flight path. The objective penalized a peak load above the buckling load and a flight speed above the flutter speed of each wing design.

In aircraft engineering, feedback control complements the structural design in keeping flexible wings within their limits. Model predictive gust load alleviation accounts for the limitations of the actuators and the system~\cite{rieck2026model}, and data-enabled policy optimization has been used to suppress flutter and alleviate loads with a single controller~\cite{wang2025unified}. Such controllers act around a given flight path, while for an AWE system, the flight path is itself the outcome of an optimization. Hence, the structural limits can be enforced where the path is generated, which complements feedback control rather than replacing it.

In this work, we incorporate the aeroelastic response of the wing into the periodic optimal control problem of a multi-megawatt fixed-wing AWES. Unlike in~\cite{fasel2021concurrent}, the wing design is fixed, and the full state and control trajectories of the pumping cycle are optimized by a gradient-based method under hard bounds on the wing deflection and twist along the whole trajectory. The contributions are:
\begin{itemize}[leftmargin=*]
    \item a modal wing model, loaded by strip-theory aerodynamics and by the inertial loads of the manoeuvring aircraft, whose quasi-static reduction keeps the overall model an index-1 differential-algebraic system;
    \item a periodic optimal control problem with bounds on wing deflection and elastic twist as path constraints;
    \item validation of the quasi-static reduction using prescribed motion and a closed-loop simulation with full modal dynamics.
\end{itemize}

\section{Model Description}\label{sec:model}
The AWES model consists of a ground station (winch), a tether, and a fixed-wing aircraft~\cite{heydarnia2025optimal,heydarnia2026airborne}. Section~\ref{sec:model_rigid} summarizes the rigid-body model for the AWES and states the changes made for this work. Section~\ref{sec:model_wing} adds the flexible-wing model.

\subsection{Rigid-body AWES model}\label{sec:model_rigid}
The winch is a drum with rotation angle, $\theta_\text{W}$, braked or driven by the generator torque, $\tau_\text{g}$. We consider the MegAWES reference aircraft~\cite{eijkelhof2022six}, modelled as a rigid body with the aerodynamic coefficient model of~\cite{heydarnia2026airborne}, which is obtained with the vortex-lattice method (VLM) in AVL~\cite{drela2025avl}. Its states are the velocity $\vectorstyle{v}^\mathcal{B}$ and angular velocity $\vectorstyle{\omega}^\mathcal{B}$ in the body frame $\mathcal{B}$, the attitude quaternion $\vectorstyle{q}$ and the position $\vectorstyle{p}^\mathcal{W}$ in the wind frame $\mathcal{W}$, and its inputs are the aileron, elevator and rudder deflections $\vectorstyle{\delta}_\text{u}$. Tether dynamics are incorporated by a quasi-static model consisting of lumped masses connected by elastic segments and loaded by gravity and aerodynamic drag~\cite{heydarnia2025optimal}. The tether configuration is computed recursively along the chain of masses, starting from the tension $T_\text{W}$, elevation $\theta_N$ and azimuth $\phi_N$ of the tether at the ground station. They are the algebraic states in~\eqref{eq:states}, and the corresponding algebraic equations require the end of the tether to coincide with the aircraft's position. With the differential states $\vectorstyle{x}$, the algebraic states $\vectorstyle{z}$ and the inputs $\vectorstyle{u}$,
\begin{equation}\label{eq:states}
    \begin{aligned}
        \vectorstyle{x} & = \big[\theta_\text{W}, \dot{\theta}_\text{W}, \vectorstyle{v}^\mathcal{B}, \vectorstyle{\omega}^\mathcal{B}, \vectorstyle{q}, \vectorstyle{p}^\mathcal{W}\big]^\top \in \mathbb{R}^{15}, \\
        \vectorstyle{z} & = \big[T_\text{W}, \theta_N, \phi_N\big]^\top \in \mathbb{R}^{3},                                                                                                                         \\
        \vectorstyle{u} & = \big[\tau_\text{g}, \vectorstyle{\delta}_\text{u}\big]^\top \in \mathbb{R}^{4},
    \end{aligned}
\end{equation}
the model is determined by a semi-explicit differential-algebraic equation (DAE)
\begin{equation}\label{eq:dae_rigid}
    \begin{aligned}
        \dot{\vectorstyle{x}} & = \vectorstyle{f}(\vectorstyle{x}, \vectorstyle{u}, \vectorstyle{z}), \\
        \vectorstyle{0}       & = \vectorstyle{g}(\vectorstyle{x}, \vectorstyle{z}),
    \end{aligned}
\end{equation}
which has differential index one, i.e., $\partial\vectorstyle{g}/\partial\vectorstyle{z}$ is nonsingular, as long as the tether remains under tension, $T_\text{W} > 0$, and does not touch the ground, $\theta_N > 0$~\cite{heydarnia2025optimal}.

Relative to our prior work~\cite{heydarnia2025optimal,heydarnia2026airborne}, two changes are made. First, the attitude is parametrized by a singularity-free unit quaternion $\vectorstyle{q}$ instead of Euler angles. Second, the aircraft position $\vectorstyle{p}^\mathcal{W}$ is given in Cartesian coordinates of the wind frame $\mathcal{W}$ instead of spherical coordinates, so that the admissible flight region becomes bounds on the states. The kinematics then read
\begin{equation}\label{eq:kinematics}
    \begin{aligned}
        \dot{\vectorstyle{q}}             & = \tfrac{1}{2}\,\vectorstyle{q} \otimes \big[0, \vectorstyle{\omega}^\mathcal{B}\big]^\top, \\
        \dot{\vectorstyle{p}}^\mathcal{W} & = \matrixstyle{R}_{\mathcal{W}\mathcal{B}}(\vectorstyle{q})\,\vectorstyle{v}^\mathcal{B},
    \end{aligned}
\end{equation}
where $\otimes$ denotes the Hamilton quaternion product and $\matrixstyle{R}_{\mathcal{W}\mathcal{B}}(\vectorstyle{q})$ the rotation from $\mathcal{B}$ to $\mathcal{W}$. The Newton--Euler equations in the body frame are unchanged and are restated because the wing model uses them,
\begin{equation}\label{eq:newton_euler}
    \begin{aligned}
        m\big(\dot{\vectorstyle{v}}^\mathcal{B} + \vectorstyle{\omega}^\mathcal{B} \times \vectorstyle{v}^\mathcal{B}\big)                              & = \vectorstyle{F}_\text{a}^\mathcal{B} + \vectorstyle{F}_\text{t}^\mathcal{B} + m\,\vectorstyle{a}_\text{g}^\mathcal{B}, \\
        \matrixstyle{J}\dot{\vectorstyle{\omega}}^\mathcal{B} + \vectorstyle{\omega}^\mathcal{B} \times \matrixstyle{J}\vectorstyle{\omega}^\mathcal{B} & = \vectorstyle{M}_\text{a}^\mathcal{B},
    \end{aligned}
\end{equation}
where $m$ and $\matrixstyle{J}$ are the mass and the inertia matrix of the aircraft, $\vectorstyle{F}_\text{a}^\mathcal{B}$ and $\vectorstyle{M}_\text{a}^\mathcal{B}$ the aerodynamic force and moment, $\vectorstyle{F}_\text{t}^\mathcal{B}(\vectorstyle{x}, \vectorstyle{z})$ the tether force, which acts at the centre of gravity of the aircraft, and $\vectorstyle{a}_\text{g}^\mathcal{B}$ the gravitational acceleration, all expressed in the body frame.

\subsection{Flexible-wing model}\label{sec:model_wing}
The wing model is assembled from established methods of flexible-aircraft modelling, namely modal reduction of a finite-element (FE) model~\cite{craig2006fundamentals}, inertial loading of the wing by the motion of the aircraft~\cite{craig1968coupling,reschke2005flight}, and strip-theory aerodynamics~\cite{wright2015introduction}. They are summarized below, together with the choices made for the AWE application.

\subsubsection{Structural modes}
The structure is represented by the FE model of the MegAWES wing of~\cite{eijkelhof2019design}, in which the wing is a single structure across the symmetry plane. All degrees of freedom of its centre section are clamped. The deformation is thus described relative to this section, which moves with the rigid-body aircraft. Let $\matrixstyle{M}, \matrixstyle{K} \in \mathbb{R}^{N_\text{d} \times N_\text{d}}$ denote the mass and stiffness matrices of the remaining $N_\text{d}$ degrees of freedom. The wing model retains the $N_\text{m}$ lowest vibration modes of this clamped structure. Their mode shapes $\matrixstyle{\Phi} \in \mathbb{R}^{N_\text{d} \times N_\text{m}}$ and natural frequencies $\matrixstyle{\Omega} = \operatorname{diag}(\omega_1, \dots, \omega_{N_\text{m}})$ follow from the generalized eigenvalue problem, with the modes normalized by the mass matrix~\cite{craig2006fundamentals}, and so that $\matrixstyle{\Phi}^\top\matrixstyle{M}\matrixstyle{\Phi} = \matrixstyle{I}$, with
\begin{equation}\label{eq:modes}
    \matrixstyle{K}\matrixstyle{\Phi} = \matrixstyle{M}\matrixstyle{\Phi}\matrixstyle{\Omega}^2.
\end{equation}

\begin{figure}[t]
    \centering
    \includegraphics[width=\columnwidth,trim={0cm 5.2cm 0cm 5cm},clip]{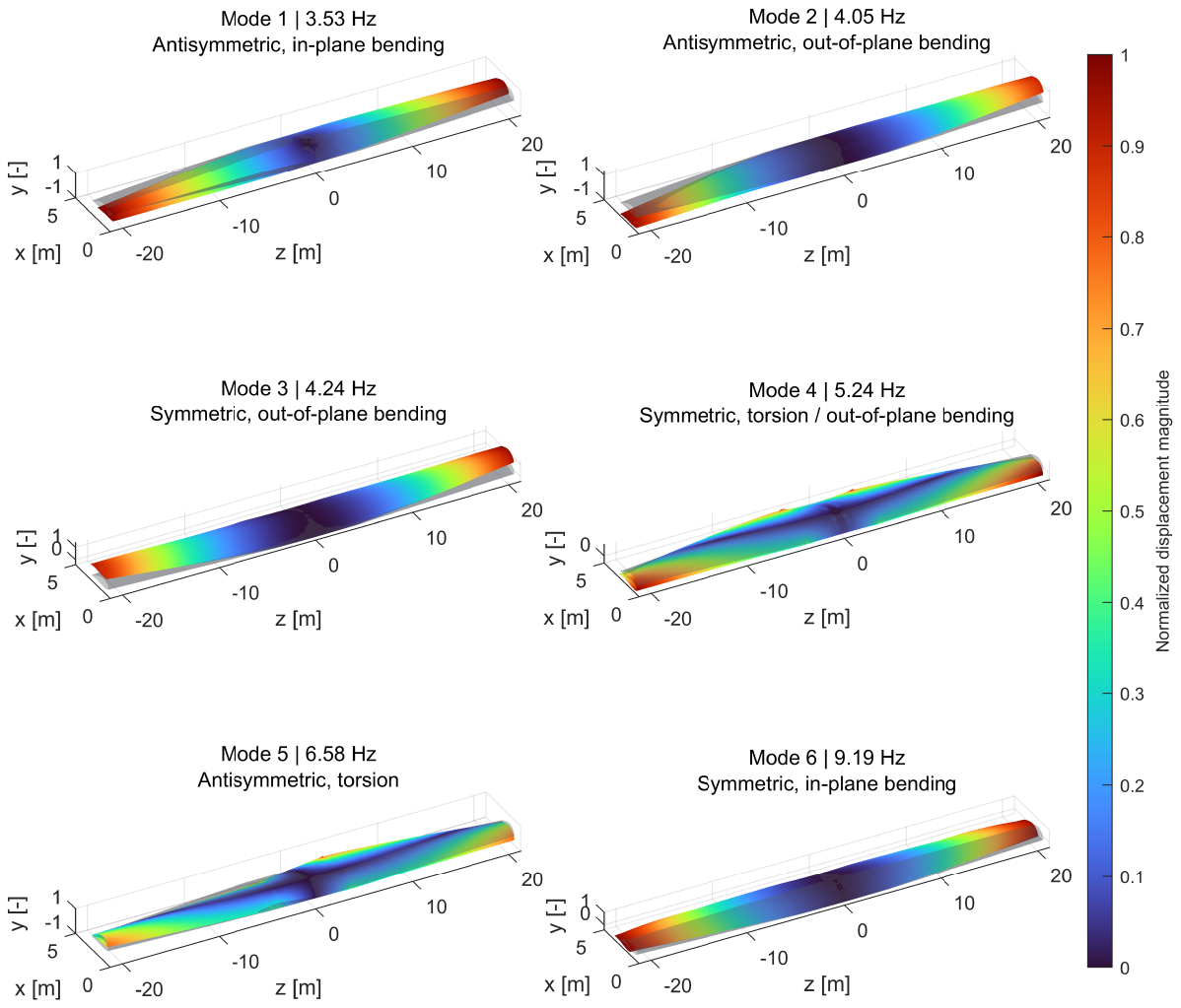}
    \caption{First six root-clamped modes of the MegAWES wing (grey indicates the undeformed shape and $x$, $y$, $z$ are the chordwise, vertical and spanwise axes of the FE frame).}
    \label{fig:modes}
\end{figure}

Fig.~\ref{fig:modes} shows the first six modes, three symmetric and three antisymmetric, with natural frequencies between \SI{3.53}{Hz} and \SI{9.19}{Hz}. The nodal velocities $\vectorstyle{v}_\text{d} \in \mathbb{R}^{N_\text{d}}$ of the wing, in the FE frame $\mathcal{F}$, follow from the rigid-body motion of the aircraft and the elastic motion relative to it,
\begin{equation}\label{eq:nodal_velocity}
    \vectorstyle{v}_\text{d} = \matrixstyle{T}_\text{r}\big[\vectorstyle{v}^\mathcal{B}, \vectorstyle{\omega}^\mathcal{B}\big]^\top + \matrixstyle{\Phi}\dot{\vectorstyle{\eta}},
\end{equation}
where $\vectorstyle{\eta} \in \mathbb{R}^{N_\text{m}}$ are the modal coordinates of the elastic displacement $\matrixstyle{\Phi}\vectorstyle{\eta}$. The matrix $\matrixstyle{T}_\text{r} \in \mathbb{R}^{N_\text{d} \times 6}$ stacks, for each FE node $n$, the block
\begin{equation}\label{eq:Tr}
    \begin{aligned}
        \matrixstyle{T}_{\text{r},n}             & =
        \begin{bmatrix}
            \matrixstyle{R}_{\mathcal{F}\mathcal{B}} & -\matrixstyle{R}_{\mathcal{F}\mathcal{B}}[\vectorstyle{r}_n]_\times \\
            \matrixstyle{0}                          & \matrixstyle{R}_{\mathcal{F}\mathcal{B}}
        \end{bmatrix}, \\
        \matrixstyle{R}_{\mathcal{F}\mathcal{B}} & =
        \begin{bmatrix}
            -1 & 0  & 0  \\
            0  & 0  & -1 \\
            0  & -1 & 0
        \end{bmatrix},
    \end{aligned}
\end{equation}
where $[\cdot]_\times$ is the cross-product matrix and $\vectorstyle{r}_n =\matrixstyle{R}_{\mathcal{F}\mathcal{B}}^\top\vectorstyle{s}_n - \vectorstyle{r}_\text{cg}$ is the position of node $n$ relative to the centre of gravity (CoG) in the body frame. Here, $\vectorstyle{s}_n$ holds the FE coordinates of the undeformed node and $\vectorstyle{r}_\text{cg}$ the position of the CoG in the body frame, both measured from the wing-root leading edge, which is the origin of the FE frame. The first block row is the velocity $\vectorstyle{v}^\mathcal{B} + \vectorstyle{\omega}^\mathcal{B} \times \vectorstyle{r}_n$ of the node, and the second its angular velocity. The rotation $\matrixstyle{R}_{\mathcal{F}\mathcal{B}}$ converts components in the body frame into components in the FE frame. The axes of the FE frame point backward, upward and along the left wing, whereas those of the body frame point forward, along the right wing and downward, hence the negative entries. Each column of $\matrixstyle{T}_\text{r}$ is thus the velocity of all nodes when the wing moves as a rigid body with a unit velocity along, or a unit angular velocity about, one axis of the body frame through the CoG.

\subsubsection{Inertial loads}
Because the modes are clamped at the centre section, the aircraft motion acts on the wing as a motion of its support~\cite{craig2006fundamentals}. Differentiating~\eqref{eq:nodal_velocity} in the rotating body frame, and neglecting the centripetal and Coriolis terms, expresses the nodal accelerations through the acceleration of the aircraft and $\ddot{\vectorstyle{\eta}}$. The elastic forces depend only on $\vectorstyle{\eta}$, because a rigid motion of the wing together with its clamped section does not strain the structure~\cite{craig1968coupling}. Substituting both into the FE equations of motion, moving the inertial and gravitational loads of the rigid motion to the right-hand side and projecting onto $\matrixstyle{\Phi}$ yields
\begin{equation}\label{eq:modal}
    \begin{aligned}
        \ddot{\vectorstyle{\eta}} + 2\matrixstyle{Z}\matrixstyle{\Omega}\dot{\vectorstyle{\eta}} + \matrixstyle{\Omega}^2\vectorstyle{\eta} & = \vectorstyle{Q}_\text{a} - \matrixstyle{\Gamma}\vectorstyle{a}_\text{r}, \\
        \matrixstyle{\Gamma}                                                                                                                & = \matrixstyle{\Phi}^\top\matrixstyle{M}\matrixstyle{T}_\text{r},
    \end{aligned}
\end{equation}
where $\vectorstyle{Q}_\text{a}$ refers to the aerodynamic generalized forces (derived in Section~\ref{sec:model_strip}). The matrix $\matrixstyle{\Gamma} \in \mathbb{R}^{N_\text{m} \times 6}$ couples the modes, through the mass matrix, to the rigid-body acceleration of the aircraft
\begin{equation}\label{eq:ar}
    \begin{aligned}
        \vectorstyle{a}_\text{r} & =
        \begin{bmatrix}
            \dot{\vectorstyle{v}}^\mathcal{B} + \vectorstyle{\omega}^\mathcal{B} \times \vectorstyle{v}^\mathcal{B} - \vectorstyle{a}_\text{g}^\mathcal{B} \\
            \dot{\vectorstyle{\omega}}^\mathcal{B}
        \end{bmatrix} \\
                                 & =
        \begin{bmatrix}
            \big(\vectorstyle{F}_\text{a}^\mathcal{B} + \vectorstyle{F}_\text{t}^\mathcal{B}\big)/m \\
            \matrixstyle{J}^{-1}\big(\vectorstyle{M}_\text{a}^\mathcal{B} - \vectorstyle{\omega}^\mathcal{B} \times \matrixstyle{J}\vectorstyle{\omega}^\mathcal{B}\big)
        \end{bmatrix},
    \end{aligned}
\end{equation}
whose first block is the specific force at the CoG, i.e., the acceleration minus the gravitational acceleration, and whose second block is the angular acceleration. From~\eqref{eq:newton_euler}, $\vectorstyle{a}_\text{r}$ depends only on $(\vectorstyle{x}, \vectorstyle{u}, \vectorstyle{z})$, and through it the wing is loaded by its own weight and by the inertia of the manoeuvre, distributed by its mass matrix. Unlike~\cite{waszak1988flight,reschke2005flight}, where the deformation of the complete aircraft is described by its free vibration modes in mean axes, so that the rigid-body acceleration does not enter the elastic equations, the present model builds on the clamped-wing FE model of~\cite{eijkelhof2019design}. It therefore uses the modes of the wing clamped at its centre section, so that the aircraft acceleration loads the wing explicitly through $\matrixstyle{\Gamma}$. With a rigid centre section, taking $\matrixstyle{T}_\text{r}$ about the CoG rather than the clamped section only changes the reference point of the rigid-body motion, and it lets $\vectorstyle{a}_\text{r}$ follow directly from~\eqref{eq:newton_euler}. Finally, the matrix $\matrixstyle{Z} = \operatorname{diag}(\zeta_j)$ in~\eqref{eq:modal} holds the structural damping ratios, and $\zeta_j = 0.01$ is assumed in the absence of measured data.

\subsubsection{Strip aerodynamics}\label{sec:model_strip}
The generalized forces $\vectorstyle{Q}_\text{a}$ follow from strip theory~\cite{wright2015introduction}. The wing is divided into $N_\text{s}$ spanwise strips with area $S_i$, chord $c_i$ and aerodynamic centre $\vectorstyle{r}_i$, taken at the quarter chord, $x_{\text{ac},i} = x_{\text{LE},i} + c_i/4$, about which the section moment coefficient is defined, with chordwise positions increasing towards the trailing edge. The section lift and moment coefficients and the induced drag coefficient of each strip are taken from the strip forces computed by AVL for the undeformed wing without control-surface deflections, and the profile drag from airfoil polars. The FE mode shapes are interpolated at the elastic axis of each strip, $x_{\text{ea},i} = x_{\text{LE},i} + 0.327c_i$, estimated from the FE model. This gives the out-of-plane, chordwise and twist components $\vectorstyle{\phi}_{z,i}^\top$, $\vectorstyle{\phi}_{x,i}^\top$ and $\vectorstyle{\phi}_{\theta,i}^\top$ of the modes, the $i$th rows of $\matrixstyle{\Phi}_z$, $\matrixstyle{\Phi}_x$ and $\matrixstyle{\Phi}_\theta \in \mathbb{R}^{N_\text{s} \times N_\text{m}}$. The air velocity and the Angle of Attack (AoA) at strip~$i$ are
\begin{equation}\label{eq:strip_kin}
    \begin{aligned}
        \vectorstyle{v}_{\text{a},i} & = \vectorstyle{v}^\mathcal{B} + \vectorstyle{\omega}^\mathcal{B} \times (\vectorstyle{r}_i - \vectorstyle{r}_\text{cg}) - \vectorstyle{v}_\text{w}^\mathcal{B} - \big[0, 0, \vectorstyle{\phi}_{z,i}^\top\dot{\vectorstyle{\eta}}\big]^\top, \\
        \alpha_i                     & = \operatorname{atan2}\big(v_{\text{a},i,3},\, v_{\text{a},i,1}\big) + \vectorstyle{\phi}_{\theta,i}^\top\vectorstyle{\eta},
    \end{aligned}
\end{equation}
where $\vectorstyle{v}_\text{w}^\mathcal{B}$ is the local wind velocity. The plunge velocity and the elastic twist of the strip thus enter the local flow velocity and angle. The loads are
\begin{equation}\label{eq:strip_loads}
    \begin{aligned}
        \bar{q}_i       & = \tfrac{1}{2}\rho\|\vectorstyle{v}_{\text{a},i}\|^2,                            \\
        L_i             & = \bar{q}_iS_i\,c_{l,i}(\alpha_i),                                               \\ D_i &= \bar{q}_iS_i\,c_{d,i}(\alpha_i),  \\
        F_{N,i}         & = L_i\cos\alpha_i + D_i\sin\alpha_i,                                             \\
        F_{T,i}         & = -L_i\sin\alpha_i + D_i\cos\alpha_i,                                            \\
        M_{\text{ea},i} & = \bar{q}_iS_ic_i\,c_{m,i}(\alpha_i) + L_i\,(x_{\text{ea},i} - x_{\text{ac},i}),
    \end{aligned}
\end{equation}
with the air density $\rho$, the normal and tangential forces $F_{N,i}$ and $F_{T,i}$, and the moment $M_{\text{ea},i}$ about the elastic axis. Their virtual work on the modes gives
\begin{equation}\label{eq:Qa}
    \vectorstyle{Q}_\text{a}(\vectorstyle{x}, \vectorstyle{\eta}, \dot{\vectorstyle{\eta}}) = \matrixstyle{\Phi}_z^\top\vectorstyle{F}_N + \matrixstyle{\Phi}_x^\top\vectorstyle{F}_T + \matrixstyle{\Phi}_\theta^\top\vectorstyle{M}_\text{ea},
\end{equation}
where $\vectorstyle{F}_N$, $\vectorstyle{F}_T$, $\vectorstyle{M}_\text{ea} \in \mathbb{R}^{N_\text{s}}$ stack the strip values. Since the twist enters $\alpha_i$, $\vectorstyle{Q}_\text{a}$ depends on $\vectorstyle{\eta}$, so the aerodynamic loads and the deformation are coupled. To obtain smooth functions for gradient-based optimization, each strip coefficient is fitted by a quadratic polynomial in $\alpha$ over $[-12^\circ, 6^\circ]$,
\begin{equation}\label{eq:poly}
    c_{\star,i}(\alpha) = a_{\star,i,2}\,\alpha^2 + a_{\star,i,1}\,\alpha + a_{\star,i,0}, \quad \star \in \{l, d, m\}.
\end{equation}

\subsubsection{Flexible AWES model}
The DAE~\eqref{eq:dae_rigid}, which represents the rigid-body dynamics, is extended with the modal coordinates and their rates,
\begin{equation}\label{eq:dae_flex}
    \begin{aligned}
        \dot{\bar{\vectorstyle{x}}} & = \bar{\vectorstyle{f}}(\bar{\vectorstyle{x}}, \vectorstyle{u}, \vectorstyle{z}) =
        \left[
            \vectorstyle{f}(\vectorstyle{x}, \vectorstyle{u}, \vectorstyle{z}),
            \dot{\vectorstyle{\eta}},
        \vectorstyle{\psi}(\bar{\vectorstyle{x}}, \vectorstyle{u}, \vectorstyle{z}) \right]^\top,                        \\
        \vectorstyle{0}             & = \vectorstyle{g}(\vectorstyle{x}, \vectorstyle{z}),
    \end{aligned}
\end{equation}
with $\bar{\vectorstyle{x}} = \big[\vectorstyle{x}, \vectorstyle{\eta}, \dot{\vectorstyle{\eta}}\big]^\top \in \mathbb{R}^{15 + 2N_\text{m}}$ and
\begin{equation}\label{eq:psi}
    \begin{aligned}
        \vectorstyle{\psi} & = \vectorstyle{Q}_\text{a}(\vectorstyle{x}, \vectorstyle{\eta}, \dot{\vectorstyle{\eta}}) - \matrixstyle{\Gamma}\vectorstyle{a}_\text{r}(\vectorstyle{x}, \vectorstyle{u}, \vectorstyle{z}) - 2\matrixstyle{Z}\matrixstyle{\Omega}\dot{\vectorstyle{\eta}} - \matrixstyle{\Omega}^2\vectorstyle{\eta}.
    \end{aligned}
\end{equation}
The extended model~\eqref{eq:dae_flex} is again a semi-explicit DAE and is referred to as the flexible AWES model. The deflection and the elastic twist at span station $y$ follow from the mode shapes,
\begin{equation}\label{eq:outputs}
    \begin{aligned}
        \delta_z(y) & = \vectorstyle{\phi}_z(y)^\top\vectorstyle{\eta}, \\ \theta_\text{e}(y) &= \vectorstyle{\phi}_\theta(y)^\top\vectorstyle{\eta},
    \end{aligned}
\end{equation}
and can be bounded in the optimal control problem.

Two aerodynamic models are used, both obtained with the VLM. The coefficient model of~\cite{heydarnia2026airborne} gives the force and moment $\vectorstyle{F}_\text{a}^\mathcal{B}$ and $\vectorstyle{M}_\text{a}^\mathcal{B}$ of the rigid aircraft in~\eqref{eq:newton_euler}, and the strip model of Section~\ref{sec:model_strip} gives the distributed wing loads in $\vectorstyle{Q}_\text{a}$. The coefficient model is used for the flight dynamics because it includes the effects of sideslip, body rates and control-surface deflections, whereas the strip coefficients depend only on the local angle of attack. During reel-out, the coefficient model gives about \SI{27}{\percent} less normal force than the strip model. The coefficient model does not account for the deformation, so the rigid-body dynamics $\vectorstyle{f}$ do not depend on $(\vectorstyle{\eta}, \dot{\vectorstyle{\eta}})$, as stated by the following assumption.

\begin{assumption}[One-way coupling]\label{ass:oneway}
    The elastic deformation of the wing does not alter the aerodynamic forces and moments of the rigid-body model~\eqref{eq:newton_euler}.
\end{assumption}
The flight state thus determines the wing loads through~\eqref{eq:ar}--\eqref{eq:Qa}, and the wing affects the trajectory only through the structural constraints of Section~\ref{sec:ocp}. Assumption~\ref{ass:oneway} is reasonable while the elastic twist, the deformation that changes the local angle of attack, remains small, which the structural constraints enforce.

\section{OPTIMAL CONTROL PROBLEM}\label{sec:ocp}

A conventional optimal control problem (OCP) for an AWES with a flexible tether and a rigid wing can be formulated as~\cite{heydarnia2025optimal}
\begin{equation}\label{eq:ocp_cont}
    \begin{aligned}
        \min_{\vectorstyle{x}(\cdot), \vectorstyle{u}(\cdot), \vectorstyle{z}(\cdot), \hat{P}, t_\text{f}} \quad & \int_0^{t_\text{f}} c(\vectorstyle{x}, \vectorstyle{u}, \vectorstyle{z}, \hat{P})\,\text{d}t + w_t\big(t_\text{f} - \hat{t}_\text{f}\big)^2 \\
        \text{s.t.} \quad
                                                                                                                 & \begin{aligned}[t]
                                                                                                                       \dot{\vectorstyle{x}} & = \vectorstyle{f}(\vectorstyle{x}, \vectorstyle{u}, \vectorstyle{z}),                           \\
                                                                                                                       \vectorstyle{0}       & = \vectorstyle{g}(\vectorstyle{x}, \vectorstyle{z}),                                            \\
                                                                                                                       \vectorstyle{0}       & \ge \vectorstyle{h}(\vectorstyle{x}, \vectorstyle{u}, \vectorstyle{z}) , t \in [0, t_\text{f}], \\
                                                                                                                       \vectorstyle{0}       & = \vectorstyle{x}(0) - \vectorstyle{x}(t_\text{f}),
                                                                                                                   \end{aligned}
    \end{aligned}
\end{equation}
where the period $t_\text{f}$ is free and kept close to the estimate $\hat{t}_\text{f}$ of the initial guess by the weight $w_t$. In~\cite{heydarnia2025optimal,heydarnia2026airborne}, this problem is solved through a homotopy of several consecutive nonlinear programs, which the wing model would make more expensive to build and solve. Here, it is solved directly, thanks to a refined initial guess.

The running cost
\begin{equation}\label{eq:cost}
    \begin{aligned}
        c = \left\|\dot{\vectorstyle{\omega}}^{\mathcal{B}}\right\|^2_{\matrixstyle{W}_\omega} + \left\|\dot{\vectorstyle{v}}^{\mathcal{B}}\right\|^2_{\matrixstyle{W}_v} + w_\beta\,\beta^2  + w_P\,\chi_\text{ro}(t)\bar{P}^2 - w_E\,P
    \end{aligned}
\end{equation}
rewards the mechanical power $P = r_\text{W}T_\text{W}\dot{\theta}_\text{W}$ at the winch with drum radius $r_\text{W}$ and penalizes the sideslip angle $\beta$ and the angular and linear accelerations of the aircraft, given by~\eqref{eq:newton_euler}, to avoid aggressive manoeuvres. The cycle is split into a reel-out phase, the first fraction $\kappa_\text{ro}$ of the period, and a retraction phase, with $\chi_\text{ro}(t) = 1$ for $t \le \kappa_\text{ro}t_\text{f}$ and $\chi_\text{ro}(t) = 0$ otherwise. During reel-out, the term $\bar{P} = P-\hat{P}$ weighted by $w_P$ keeps the power close to the reference $\hat{P}$, a free decision variable, and thereby smooths the generated power in this phase. Here, $\matrixstyle{W}_\omega\succ0$, $\matrixstyle{W}_v\succ0$, $w_\beta>0$, $w_P>0$ and $w_E>0$ are user-defined weights.

Problem~\eqref{eq:ocp_cont} is converted into a nonlinear program by a direct transcription method on a uniform grid of $N$ intervals of length $\Delta t = t_\text{f}/N$, with the controls constant on each interval. The dynamics are imposed with the implicit midpoint rule instead of the implicit Euler rule of~\cite{heydarnia2025optimal}. The midpoint rule is the one-stage Gauss--Legendre method, the cheapest method of this family, and it is second-order accurate, A-stable and free of numerical dissipation. A cheap scheme allows the largest number of grid nodes for a given computational effort, which matters here because the wing loads must be resolved along the whole cycle.

For the quaternion, the midpoint rule becomes the Cayley scheme
\begin{equation}\label{eq:cayley}
    \begin{aligned}
        \vectorstyle{q}_{k+1} - \vectorstyle{q}_k & = \tfrac{\Delta t}{4}\,\big(\vectorstyle{q}_k + \vectorstyle{q}_{k+1}\big) \otimes \big[0, \vectorstyle{\omega}^\mathcal{B}_{k+\frac{1}{2}}\big]^\top,
    \end{aligned}
\end{equation}
where the subscript $k+\frac{1}{2}$ denotes the mean of two neighbouring nodes. Since right-multiplication by a pure quaternion is a skew-symmetric linear map, the norm is preserved exactly. Consequently, we have that $\|\vectorstyle{q}_{k+1}\| = \|\vectorstyle{q}_k\|$, and a single constraint $\|\vectorstyle{q}_0\| = 1$ at the first node keeps the quaternion unit. A norm constraint at every node would be redundant and violate the linear independence constraint qualification (LICQ). For the same reason, periodicity is imposed on all other states through the selection matrix $\matrixstyle{\Pi}$ and, for the attitude, only on the vector part of the relative quaternion,
\begin{equation}\label{eq:periodic}
    \begin{aligned}
        \vectorstyle{0} & = \matrixstyle{\Pi}\big(\vectorstyle{x}_N - \vectorstyle{x}_0\big),            \\
        \vectorstyle{0} & = \operatorname{vec}\big(\vectorstyle{q}_0^{*} \otimes \vectorstyle{q}_N\big),
    \end{aligned}
\end{equation}
where $\vectorstyle{q}^*$ is the conjugate quaternion. The second condition states that the rotation from the first to the last attitude has no vector part. Together with the preserved norm, it gives $\vectorstyle{q}_N = \pm\vectorstyle{q}_0$, the same attitude. This is the projection method of~\cite{gros2018numerical} for periodic OCPs whose dynamics hold invariants, here applied to the unit quaternion, and it recovers the LICQ.

The tether enters through the algebraic states. At every node, $\vectorstyle{z}_k$ is a decision variable and the quasi-static tether equations $\vectorstyle{0}=\vectorstyle{g}(\vectorstyle{x}_k, \vectorstyle{z}_k)$, which place the end of the tether at the aircraft, are imposed as constraints \cite{heydarnia2025optimal}. The tether equilibrium is thus solved together with the dynamics, and at the first node it yields consistent initial conditions. In the discretized dynamics, the tether force uses the mean $\vectorstyle{z}_{k+\frac{1}{2}}$.

The path constraints $\vectorstyle{h}$ include bounds on quantities such as the airspeed, the aerodynamic angles, the angles between the tether and the aircraft~\cite{heydarnia2025optimal}, the power and the accelerations of the aircraft and the winch, and they keep the reel-out speed non-negative during the reel-out phase. Several of these limits, for instance the maximum power, the maximum acceleration and the maximum airspeed, are usually chosen by intuition, in particular for an aircraft of the size of MegAWES, which has not been built. They protect the structure only indirectly. Constraints that act directly on the aeroelastic response of the wing are therefore important, and the OCP is extended accordingly in the following.

Replacing~\eqref{eq:dae_rigid} by the flexible AWES model~\eqref{eq:dae_flex} in~\eqref{eq:ocp_cont} is too expensive, even with the midpoint rule, because resolving the elastic modes requires time steps well below the period of the first elastic mode, much finer than the grid that resolves the flight dynamics. The Optimal Control Problem therefore relies on the following assumption.
\begin{assumption}[Quasi-static wing]\label{ass:qs}
    Along the planned trajectories, the inertia and damping forces of the elastic motion are negligible compared with the elastic forces $\matrixstyle{\Omega}^2\vectorstyle{\eta}$.
\end{assumption}
Setting $\ddot{\vectorstyle{\eta}} = \dot{\vectorstyle{\eta}} = \vectorstyle{0}$ in~\eqref{eq:dae_flex} moves $\vectorstyle{\eta}$ from the differential to the algebraic part,
\begin{equation}\label{eq:dae_qs}
    \begin{aligned}
        \dot{\vectorstyle{x}} & = \vectorstyle{f}(\vectorstyle{x}, \vectorstyle{u}, \vectorstyle{z}),                                                                                                                              \\
        \vectorstyle{0}       & = \bar{\vectorstyle{g}}(\vectorstyle{x}, \vectorstyle{u}, \bar{\vectorstyle{z}}) =
        \begin{bmatrix}
            \vectorstyle{g}(\vectorstyle{x}, \vectorstyle{z}) \\
            \matrixstyle{\Omega}^2\vectorstyle{\eta} - \vectorstyle{Q}_\text{a}(\vectorstyle{x}, \vectorstyle{\eta}, \vectorstyle{0}) + \matrixstyle{\Gamma}\vectorstyle{a}_\text{r}(\vectorstyle{x}, \vectorstyle{u}, \vectorstyle{z})
        \end{bmatrix}, \\
        \bar{\vectorstyle{z}} & = \big[\vectorstyle{z}, \vectorstyle{\eta}\big]^\top.
    \end{aligned}
\end{equation}

The reduced model~\eqref{eq:dae_qs} remains an index-1 DAE, because its algebraic Jacobian
\begin{equation}\label{eq:jacobian}
    \frac{\partial\bar{\vectorstyle{g}}}{\partial\bar{\vectorstyle{z}}} =
    \begin{bmatrix}
        \dfrac{\partial\vectorstyle{g}}{\partial\vectorstyle{z}}                              & \matrixstyle{0}                                                                               \\[1ex]
        \matrixstyle{\Gamma}\dfrac{\partial\vectorstyle{a}_\text{r}}{\partial\vectorstyle{z}} & \matrixstyle{\Omega}^2 - \dfrac{\partial\vectorstyle{Q}_\text{a}}{\partial\vectorstyle{\eta}}
    \end{bmatrix}
\end{equation}
is block lower-triangular, $\partial\vectorstyle{g}/\partial\vectorstyle{z}$ is nonsingular under the conditions stated after~\eqref{eq:dae_rigid}, and $\matrixstyle{\Omega}^2 - \partial\vectorstyle{Q}_\text{a}/\partial\vectorstyle{\eta}$ is nonsingular away from static aeroelastic divergence. Unlike in~\eqref{eq:dae_rigid}, the algebraic equations depend on $\vectorstyle{u}$, through $\vectorstyle{a}_\text{r}$.

Under the quasi-static approximation, the wing is at every instant in static aeroelastic equilibrium with the current aerodynamic and inertial loads. Only the inertia and damping of the elastic motion are neglected, while the inertial loads $\matrixstyle{\Gamma}\vectorstyle{a}_\text{r}$ of the manoeuvring aircraft are kept. Both this equilibrium and the index-1 property rely on the matrix $\matrixstyle{\Omega}^2 - \partial\vectorstyle{Q}_\text{a}/\partial\vectorstyle{\eta}$ in~\eqref{eq:jacobian}, which is the structural stiffness of the wing modified by an aerodynamic stiffness that arises because the elastic twist enters the angle of attack in~\eqref{eq:strip_kin}. Since the aerodynamic centres lie ahead of the elastic axis, the moment of the lift about this axis twists the wing nose-up, the twist raises the angle of attack and hence the lift, and the larger lift twists the wing further. The torsional stiffness of the structure resists this loop, whereas the aerodynamic stiffness grows with the dynamic pressure. At the divergence speed the two cancel, the matrix becomes singular, and the wing has no equilibrium, so that its twist grows without bound~\cite{palacios2023dynamics}. Below this speed, this loop only softens the torsional response of the wing. Divergence is a static instability, unlike flutter, which is oscillatory and involves the elastic dynamics neglected in~\eqref{eq:dae_qs}. Besides the existence of the equilibrium, Assumption~\ref{ass:qs} requires the loads to vary slowly compared with the natural frequency $\omega_1$~\cite{palacios2023dynamics}.

To quantify how fast the loads vary along a periodic trajectory relative to the first natural frequency, the loads can be expanded in a harmonic series with angular frequencies $\nu = \ell\,(2\pi/t_\text{f})$, $\ell \in \mathbb{N}$. Neglecting damping, each harmonic scales the static response of the first mode by $A(\nu) = 1/\vert{}1 - (\nu/\omega_1)^2\vert{}$~\cite[Sec.~4.1]{craig2006fundamentals}, so the relative error of the quasi-static wing is about $(\nu/\omega_1)^2$ for $\nu \ll \omega_1$. Because the trajectory is itself the outcome of the optimization, Assumption~\ref{ass:qs} cannot be verified beforehand. It is therefore checked a posteriori by simulations that keep the modal dynamics, with the aircraft motion either prescribed by the optimized trajectory or simulated in closed loop (Section~\ref{sec:results}).

The OCP with the elastic wing is~\eqref{eq:ocp_cont} with~\eqref{eq:dae_rigid} replaced by~\eqref{eq:dae_qs} and with bounds on the deflection and twist~\eqref{eq:outputs} added to the path constraints. In the transcription, one set of modal coordinates $\vectorstyle{\eta}_k$ is introduced per interval and the wing equilibrium is imposed at the midpoint, where the dynamics evaluate the aerodynamics and the accelerations,
\begin{equation}\label{eq:wing_k}
    \vectorstyle{0} = \matrixstyle{\Omega}^2\vectorstyle{\eta}_k - \vectorstyle{Q}_\text{a}\big(\vectorstyle{x}_{k+\frac{1}{2}}, \vectorstyle{\eta}_k, \vectorstyle{0}\big) + \matrixstyle{\Gamma}\vectorstyle{a}_\text{r}\big(\vectorstyle{x}_{k+\frac{1}{2}}, \vectorstyle{u}_k, \vectorstyle{z}_{k+\frac{1}{2}}\big),
\end{equation}
and the deflection $\delta_z$ and twist $\theta_\text{e}$ of~\eqref{eq:outputs} are bounded at a few selected span stations $y_m$ on both wings,
\begin{equation}\label{eq:struct}
    \big|\vectorstyle{\phi}_z(y_m)^\top\vectorstyle{\eta}_k\big| \le \bar{\delta}_z, \quad \big|\vectorstyle{\phi}_\theta(y_m)^\top\vectorstyle{\eta}_k\big| \le \bar{\theta}_\text{e}.
\end{equation}

To limit the size of the nonlinear program, the aerodynamic loads $\vectorstyle{Q}_\text{a}$ in~\eqref{eq:wing_k} are evaluated on $N_\text{p} < N_\text{s}$ panels, each of which merges neighbouring strips of one semispan. Each panel is treated as one wide strip with the total area of its strips, and its coefficients and mode shapes are weighted averages over these strips. When all strips of a panel see the same angle of attack, the panel carries the same aerodynamic load as its strips together. The simplification is that all strips of a panel see the same inflow and the same elastic twist, and that the panel load acts on the modes through the mode shapes averaged over its strips. The resulting error is small, as shown in Section~\ref{sec:results}. The merging affects only the loads, and the deflection and twist in~\eqref{eq:struct} are still taken from the mode shapes at the span stations $y_m$ themselves.

\section{RESULTS}\label{sec:results}

The method is applied to the MegAWES aircraft~\cite{eijkelhof2022six} in a logarithmic wind profile with a wind speed of \SI{15}{m/s} at the reference height of \SI{250}{m}. The optimized cycle contains six loops during reel-out, and the grid has $N = 266$ intervals of about \SI{0.47}{s}. The wing model uses the $N_\text{m} = 6$ modes of Fig.~\ref{fig:modes} and $N_\text{p} = 20$ aerodynamic panels, which merge the $N_\text{s} = 100$ strips. The span stations $y_m$ in~\eqref{eq:struct} lie at \SI{50}{\percent} and \SI{99}{\percent} of the semispan of both wings. The deflection limit is $\bar{\delta}_z = \SI{1}{m}$, and the twist limit $\bar{\theta}_\text{e}$ is varied between $0.6^\circ$ and $1^\circ$. To isolate the effect of the wing limits, the OCP is also solved with a rigid wing, i.e., without~\eqref{eq:wing_k} and~\eqref{eq:struct}, and the quasi-static wing is evaluated afterwards along this rigid-wing trajectory. The limits are illustrative rather than structurally motivated. The deflection limit lies just below the peak deflection of the rigid-wing trajectory, and the twist limit is varied to show the trade-off with the power. The nonlinear programs are solved with IPOPT~\cite{wachter2006implementation} through CasADi~\cite{andersson2019casadi}. Table~\ref{tab:sweep} lists their solve times on an Intel Core i7-1265U processor.



Fig.~\ref{fig:traj} shows the optimized trajectories for the rigid wing and for the elastic wing with $\bar{\theta}_\text{e} = 0.6^\circ$, which is the main case of this section. The loops of both cover almost the same region, at a mean elevation of about $28^\circ$. The retraction paths differ more, since the elastic trajectory leaves the last loop at another point. The retraction costs only about \SI{2.4}{\percent} of the reel-out energy in both cases, so its path has little influence on the objective.

\subsection{Cost of the wing limits}\label{sec:results_sweep}
\begin{figure}[t]
    \centering
    \includegraphics[width=.9\columnwidth]{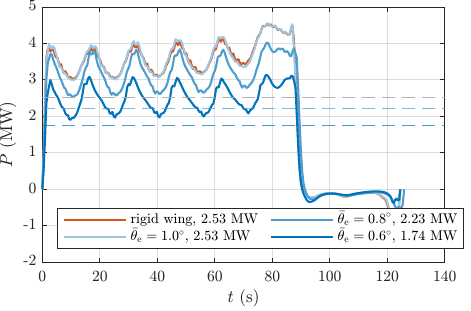}
    \caption{Power over the cycle for the rigid-wing trajectory and for twist limits $\bar{\theta}_\text{e}$ between $0.6^\circ$ and $1^\circ$, with $\bar{\delta}_z = \SI{1}{m}$. The dashed lines and the values in the legend give the mean powers.}
    \label{fig:sweep}

    \centering
    \includegraphics[width=.9\columnwidth]{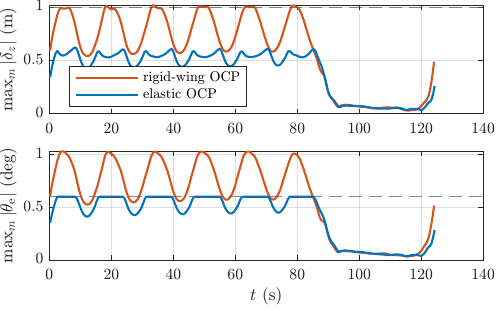}
    \caption{Largest deflection (top) and elastic twist (bottom) over the span stations along the cycle, for the rigid-wing trajectory and the elastic trajectory with $\bar{\theta}_\text{e} = 0.6^\circ$. The dashed lines mark the limits.}
    \label{fig:wing}
\end{figure}

Table~\ref{tab:sweep} compares the rigid-wing trajectory with the elastic trajectories. Along the rigid-wing trajectory, the quasi-static wing reaches a twist of $1.03^\circ$ and a deflection of \SI{1.02}{m}. At $\bar{\theta}_\text{e} = 1^\circ$, the limits thus only just bind and cost no measurable power. Tightening the twist limit to $0.8^\circ$ and $0.6^\circ$ costs \SI{12}{\percent} and \SI{31}{\percent} of the mean power, and the peak tether tension falls almost in proportion to the twist limit. Fig.~\ref{fig:sweep} shows that the power is lost during reel-out, while the retraction hardly changes.

\begin{table}[b]
    \caption{Comparative results for three twist limits, with $\bar{\delta}_z = \SI{1}{m}$}
    \label{tab:sweep}
    \centering
    \footnotesize
    \begin{tabular}{lcccc}
        \toprule
                                         & Rigid & \multicolumn{3}{c}{Elastic, $\bar{\theta}_\text{e}$}                             \\
        \cmidrule(lr){3-5}
                                         & wing  & $1^\circ$                                            & $0.8^\circ$ & $0.6^\circ$ \\
        \midrule
        Mean power (MW)                  & 2.53  & 2.53                                                 & 2.23        & 1.74        \\
        Change in mean power (\%)        & --    & $+0.1$                                               & $-11.9$     & $-31.1$     \\
        Period (s)                       & 124.4 & 124.4                                                & 125.8       & 124.6       \\
        Peak tether tension (MN)         & 1.50  & 1.50                                                 & 1.22        & 0.89        \\
        Peak twist (deg)                 & 1.03  & 1.00                                                 & 0.80        & 0.60        \\
        Peak deflection (m)              & 1.02  & 1.00                                                 & 0.82        & 0.62        \\
        Twist limit active (\% of cycle) & --    & 9.4                                                  & 25.6        & 37.2        \\
        Solve time (s)                   & 217   & 295                                                  & 351         & 253         \\
        \bottomrule
    \end{tabular}
    \vspace{12pt}

    \caption{Forward simulation for $\bar{\theta}_\text{e} = 0.6^\circ$ (in percent of the limits)}
    \label{tab:check}
    \centering
    \footnotesize
    \begin{tabular}{lcc}
        \toprule
                                                                  & Deflection & Twist \\
        \midrule
        Error of the OCP wing model (fits, panels)                & 0.13       & 0.23  \\
        \multicolumn{3}{l}{Error of the quasi-static assumption}                       \\
        \quad prescribed motion, whole cycle                      & 1.0        & 1.8   \\
        \quad prescribed motion, within \SI{5}{\percent} of limit & --         & 1.2   \\
        \quad closed loop, after \SI{2}{s}                        & 0.3        & 0.6   \\
        \multicolumn{3}{l}{Largest exceedance of the limit}                            \\
        \quad prescribed motion                                   & --         & 1.2   \\
        \quad closed loop                                         & --         & 1.2   \\
        Closed loop, time above the limit (\% of cycle)           & 0          & 12    \\
        \bottomrule
    \end{tabular}
\end{table}

For the main case of $\bar{\theta}_\text{e} = 0.6^\circ$, Fig.~\ref{fig:wing} shows that the twist limit binds at the outer station of one wing, where the wing twists nose-down. This is mainly due to the lift, which bends the wing upward through out-of-plane modes that also twist it nose-down ($-0.54^\circ$). This outweighs the nose-up moment of the lift about the elastic axis ($+0.05^\circ$) discussed in Section~\ref{sec:ocp}. Instead of a twist peak once per loop, the elastic trajectory holds the twist at the limit over a large part of every loop. The deflection stays at \SI{62}{\percent} of its limit, since limiting the twist also limits the deflection. To meet the twist limit, the trajectory lowers the aerodynamic loading during reel-out. The mean angle of attack drops from $3.7^\circ$ to $1.3^\circ$, and the mean airspeed from 102 to \SI{94}{m/s}.

\subsection{Validity of the quasi-static wing}\label{sec:results_qs}
Along the optimized trajectory, \SI{99}{\percent} of the variation of each modal coordinate lies below \SI{7}{\percent} of the natural frequency of that mode. For these harmonics, $A(\nu)$ is at most 1.005, so that the quasi-static wing underestimates the dynamic response by less than \SI{0.5}{\percent}. The optimized trajectory cannot represent faster variations, because its intervals of \SI{0.47}{s} resolve frequencies only up to the Nyquist frequency of $1/(2 \times \SI{0.47}{s}) \approx \SI{1.07}{Hz}$. The quasi-static wing is therefore checked by two simulations in which the wing keeps the inertia and damping of~\eqref{eq:dae_flex}, with a time step of \SI{0.01}{s}. With prescribed motion, the aircraft motion is taken from the optimized trajectory, which Assumption~\ref{ass:oneway} permits, and only the modal equations are integrated. The whole pumping cycle is simulated twice in a row, and only the second repetition is evaluated, where the start-up transient of the wing due to its initialization has decayed. In closed loop, the full flexible AWES model~\eqref{eq:dae_flex} is simulated, and the controller of~\cite{heydarnia2026airborne} tracks the optimized trajectory with small deviations. Both simulations use all $N_\text{s}$ strips and interpolate the AVL strip data by B-splines instead of the quadratic fits~\eqref{eq:poly}. Table~\ref{tab:check} summarizes the results.
The first row of Table~\ref{tab:check} shows that the quadratic fits and the panel merging of Section~\ref{sec:ocp} hardly affect the wing response. The next rows compare the dynamic wing with the quasi-static wing of the same simulation model. With prescribed motion, neglecting the wing dynamics changes the deflection and twist by at most \SI{1.8}{\percent} of the limits, and by at most \SI{1.2}{\percent} where the wing is near its limit. This exceeds the estimate based on $A(\nu)$, which misses the fast load changes at the start of reel-out. In closed loop, the aircraft stays within \SI{3.8}{m} of the optimized trajectory at the same mean power. By Assumption~\ref{ass:oneway}, the wing does not affect this tracking, so the closed loop shows how the wing responds when the flight deviates from the optimized trajectory. After the first two seconds, in which the simulated wing settles from its initial state, the error of the quasi-static wing is even smaller than with prescribed motion. The last rows show that only the twist exceeds its limit, by at most \SI{1.2}{\percent}, and only where the optimized trajectory itself is at the limit, because it meets the limit without margin. Assumption~\ref{ass:qs} thus holds for the MegAWES wing, whose first natural frequency lies well above the load frequencies of the trajectory, both along the optimized trajectory and in closed loop.

\section{CONCLUSIONS}\label{sec:conclusions}
This paper has included the aeroelastic response of the wing in the periodic OCP of a pumping-cycle fixed-wing AWE system. For the MegAWES aircraft, the twist limit binds at the outer span station during the high-load part of the loops. Limiting the twist to $0.8^\circ$ and $0.6^\circ$ costs \SI{12}{\percent} and \SI{31}{\percent} of the mean power, since the trajectory lowers the aerodynamic loading and hence the tether tension during reel-out. Rigid-wing trajectory optimization can overestimate the power that the wing limits permit. Simulations with the dynamic modal model confirm the quasi-static reduction to within \SI{2}{\percent} of the limits. The structural limits of the wing can thus be enforced where the trajectory is generated, which complements load-alleviating feedback control.

These results rest on a simplified wing model. It retains six modes and takes its strip aerodynamics from a different AVL computation than the rigid-body model. Since the strip model gives more lift, the predicted wing response is conservative in this respect, although retaining more modes would raise the twist. Future work will extend the wing model accordingly and replace the limits by allowables from a structural analysis of the wing. Deriving the rigid-body and the strip aerodynamics from one model would also allow a two-way coupling. Furthermore, unsteady loads, such as gusts, call for the dynamic wing model in feedback control or in robust trajectory planning.

\addtolength{\textheight}{-2cm}   





\section*{ACKNOWLEDGMENT}

The authors would like to acknowledge the support of the Energy Transition Fund of FPS Economy, project BORNE (\textit{Belgian Offshore
    aiRborne wiNd Energy}). The authors would also like to thank Dylan Eijkelhof for sharing his modal dynamics analysis results


\small
\bibliographystyle{IEEEtran}
\bibliography{references}

\end{document}